\documentclass[twocolumn]{aastex6}
\usepackage{amsmath,amssymb,lmodern,mathrsfs,ulem,mathtools}
\usepackage{graphicx}
\usepackage{natbib}
\usepackage{xcolor}
\usepackage{import}

\usepackage{lipsum}

\definecolor{forest}{RGB}{34,139,34}
\definecolor{lime}{RGB}{50,205,50}

\slugcomment{Prepared for submission to The Astrophysical Journal \\ ~}
\watermark{DRAFT V1.0}
\setwatermarkfontsize{2cm} 

\begin{document}

\title{magnetic structures at the boundary of the closed corona: \\ interpretation of S-web arcs}

\author{Roger B. Scott}
\author{David I. Pontin}
\affiliation{School of Science and Engineering \\ University of Dundee \\ Dundee, DD1 4HN, Scotland}

\author{Anthony R. Yeates}
\author{Peter F. Wyper}
\affiliation{Department of Mathematical Sciences \\ Durham University \\ Durham, DH1 3LE, UK}

\date{\today}
%\submitjournal{The Astrophysical Journal}
%\received{-- ~}
%\revised{-- ~}
%\accepted{-- ~}
%\published{--}

\begin{abstract}
The topology of magnetic fields near the open-closed flux boundary in the Sun's corona is an important influencing factor in the process of interchange reconnection, whereby plasma is exchanged between open and closed flux domains. 
Maps of the magnetic squashing factor at the radial outer boundary in coronal field models reveal the presence of the so-called `S-web', and suggest that interchange reconnection could potentially deposit closed coronal material into high-latitude regions far from the heliospheric current sheet. 
Here we demonstrate that certain features of the S-web reveal the underlying topological structure of the magnetic field. 
Specifically, in order for the arcing bands of highly squashed magnetic flux of the S-web to terminate or intersect away from the helmet streamer apex, there must be a null spine line that maps a finite segment of the photospheric open-closed boundary up to a singular point in the open flux domain. 
We propose that this association between null spine lines and arc termination points may be used to identify locations in the heliosphere that are preferential for the appearance of solar energetic particles or slow solar wind plasma with certain characteristics.
\end{abstract}

\maketitle

\section{Introduction}\label{introduction.sec}

The structuring of the open-closed boundary (OCB) in coronal magnetic fields is very important to the process of interchange reconnection \citep[IR,][]{Crooker:2002}, whereby plasma is exchanged between open and closed flux domains. 
The connection between interchange reconnection and active region upflows has been explored by \cite{vandriel-gesztelyi:281.237}, who used spectroscopic observations from the {\it Hinode EUV Imaging Spectrometer} (EIS) to infer plasma upflow velocities, which they associated with quasi-separatrix layers \citep[see also][]{Baker:2017.292.46}. 
This is consistent with the study of \cite{Brooks:2015}, who concluded that the majority of the slow solar wind (SSW) originates near the boundaries of active regions. 
These results support the proposition of \cite{Antiochos:2011.732} that reconnection within narrow corridors of open flux could explain the latitudinal extent of the slow solar wind -- they showed that these narrow photospheric open flux corridors are associated with an array of separatrices and quasi-separatrix layers (QSLs) in the coronal magnetic field which they termed the ``S-web" \citep[see also][]{Crooker:2012, Higginson:2017b}. 

In global models some of these structures are associated with observed pseudo-streamers \citep{Wang:2007}, and there is growing evidence that these are linked to the SSW outflow \citep{Owens:2013}. 
Statistical studies of the S-web have been carried out, but these have focussed largely on solar cycle dependence. 
\cite{Owens:2014} measured the typical distribution of dipolar streamers and pseudo-streamers and showed that the latitudinal extent of the pseudo-streamer belt increases with sunspot number. 
Similarly, \cite{Fujiki:2016} showed that the photospheric pattern of coronal hole footprints tracks closely with butterfly diagrams of the line-of-sight magnetic field. 
These results support the conclusion that active region dynamics could contribute to the slow solar wind; however, as these studies span multiple solar cycles, they must rely on computationally inexpensive measures of the magnetic topology, and offer little insight into the structuring of the underlying magnetic field.

Reconnection at the OCB may also be important in understanding solar energetic particle (SEP) events. 
Many of these SEP events are proposed to have a flare-accelerated component \citep[e.g.][]{Li:2009,Masson:2009.257,McCracken:2012}, and while the flare itself takes place in the closed corona, interchange reconnection is necessary to permit the accelerated particles to access open magnetic field lines and thus escape into the heliosphere, as described by \cite{Masson:2013}.

The magnetic topology of the coronal magnetic field is characterised by a complicated array of magnetic null points, their associated spine lines and separatrix surfaces \citep[e.g.][]{Longcope:2009.soph,Platten:2014j,Freed:2015}. 
\cite{Platten:2014j} described various different structures that can form from combinations of separatrix surfaces, many of these being pertinent to the OCB structure. 
\cite{Titov:2011.731} considered the topology of the OCB using an analytical model of an active region that forms a quasi-separatrix layer with a non-trivial intersection with the separatrix surface of the global helmet streamer \citep[GHS, also called the dipole streamer in][]{Owens:2014}. 
They showed how a narrow open flux corridor with a quasi-separatrix layer \citep[QSL,][]{Priest:1995dc} formed on a hyperbolic flux tube \cite[HFT,][]{Titov:2002bv} could be transformed continuously into a genuine separatrix surface as the corridor narrows to zero width, after which a new coronal null is formed. 

The study of \cite{Titov:2011.731} also highlights the importance of a wholistic approach to studies of the OCB, which treats null points, separatrix surfaces and HFTs simultaneously. 
In our current investigation we endeavour to reconcile the topological and geometrical structures from empirical models with those observed in global magnetic field extrapolations. 
This is motivated by a need to improve our ability to recognise where these structures occur in magnetic field models from their signatures on the outermost imprint of the S-web so that, together with studies of dynamic interchange reconnection processes, we can improve our ability to identify locations in the heliosophere that are likely to host plasma and SEPs with certain closed-corona characteristics. 

The structure of this investigation is as follows. 
In section \ref{glob_mod.sec} we describe our magnetic model and characterisation methods. 
Then, in section \ref{obs_hqa.sec} we highlight some observed patterns in the formation of structures within the model. 
In section \ref{emp_hqa.sec} we detail some distinctions between structures formed with and without magnetic nulls present, and how these relate to the observed formation patterns. 
Finally, in section \ref{disc.sec} we conclude with a discussion of implications and future research.

\section{Global Field Model}\label{glob_mod.sec}

\subsection{PFSS Magnetic Field}

For the magnetic field model we begin with GONG magnetogram data, which provides a synoptic measure of the full-sun radial magnetic field at a resolution of $180 \times 360$ in solar sine-latitude and Carrington longitude, respectively.  
From the radial source data, we compute the potential field source surface (PFSS) magnetic field using a finite-difference method similar to \cite{vanBallegooijen:2000}. 
A Python implementation of this method is available on \texttt{github}\footnote{\url{https://github.com/antyeates1983/pfss}}, where a full description is available. 
The data is smoothed by multiplying the coefficients of the spherical harmonic expansion (with harmonic degree $l$) by a Gaussian filter of the form $f = \exp(-l (l+1) k)$, with $k=0.002$, so that the filter amplitude falls to one half at a harmonic degree of $l_{\rm f} \sim 18$. 
This is equivalent to allowing the source magnetogram data to diffuse for a time of $0.002$ of the global diffusion time. 
Here we place the outer boundary at 2.5 solar radii, and we choose a resolution of $61 \times 180 \times 360$ in log-radius, sine-latitude, and longitude, respectively. 
For this investigation we focus on data from 29 July, 2014; however, we have constructed ten additional datasets, each taken on 01 January, every year from 2008 -- 2017, for a total of eleven independent magnetic models, which will be detailed in an expanded future investigation. 
Despite variations in the details of each model, the structures present in the 29th July 2014 data set have been found to be representative of those found in the other ten datasets.

\begin{figure}[h]
\center
\includegraphics[width=\linewidth]{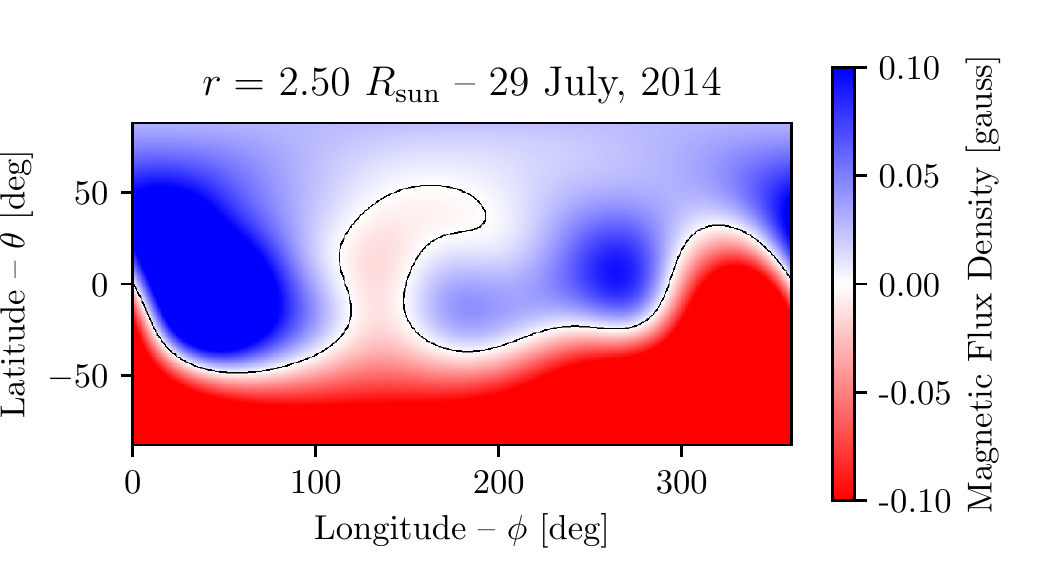}
\includegraphics[width=\linewidth]{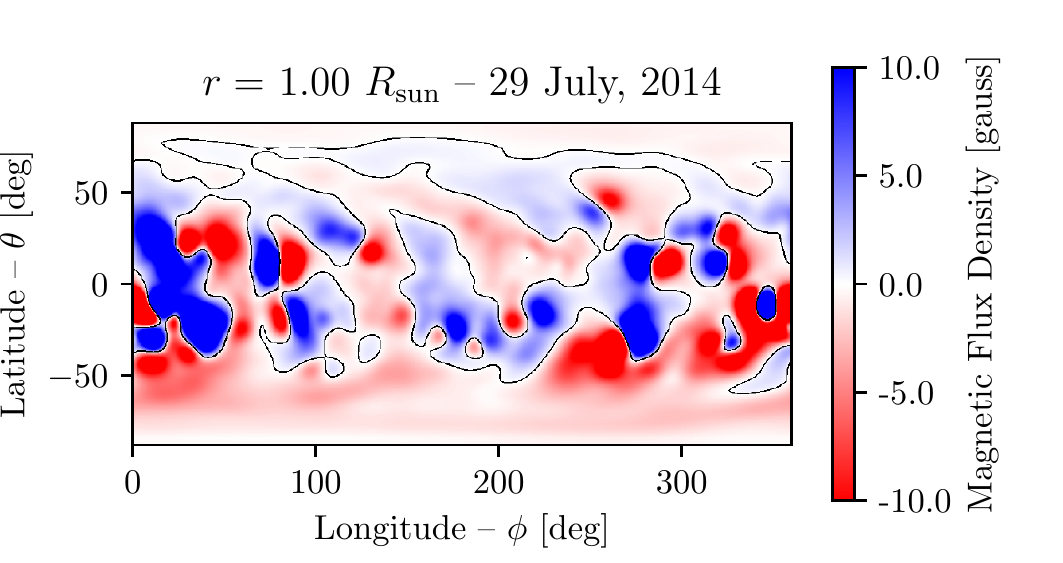}
\caption{The PFSS radial magnetic field from 29 July 2014 GONG magnetogram data, with the polarity inversion line indicated in black. The outermost boundary (source surface) is shown in the upper panel, while the innermost boundary (photosphere) is depicted below.  In the upper panel, the SSPIL divides the positive and negative flux domains, which coincide with the predominantly unipolar northern and southern hemispheres.}\label{source_field.fig}
\end{figure}

In Figure \ref{source_field.fig} the radial component of our model magnetic field is shown at the inner and outer radial boundaries, resampled to a resolution of $480 \times 960$ in Colatitude and Longitude for consistency with the analysis described in section \ref{squash.sec}. 
The source surface polarity inversion line (SSPIL) is depicted by the meandering black line in the upper panel, between the predominantly red and blue hemispheres. 
The SSPIL is the base of the heliospheric current sheet (HCS) in extended coronal models. 
In our PFSS model the HCS is a null-line (since $B_\theta=B_\phi=0$ at the source surface by construction), from which emanates a pair of separatrix surfaces that define the GHS and separate the open and closed magnetic flux domains.
It is worth noting that, this model being taken very near the pole reversal at the maximum of solar cycle 24 \citep{Gopalswamy:2016m}, the northern polar region is dominated by negative flux, opposite the global dipole field.

\subsection{Squashing Degree Estimate}\label{squash.sec}

In this investigation we use the perpendicular magnetic squashing factor, $Q_\perp$, \citep{Titov:2007gn, Pariat:2012} to identify regions of high complexity in the magnetic field line mapping, which we call ``high-Q volumes'' (HQVs). 
This method does not distinguish QSLs from genuine topological structures, and does not explicitly determine the magnetic skeleton as would, for example, the method of \cite{Haynes:2010}. 
However, it has the advantage that separatrix surfaces and quasi separatrix layers are identified simultaneously, each being potentially important to the process of interchange reconnection. 

\begin{figure*}
\center
\includegraphics[width=\linewidth]{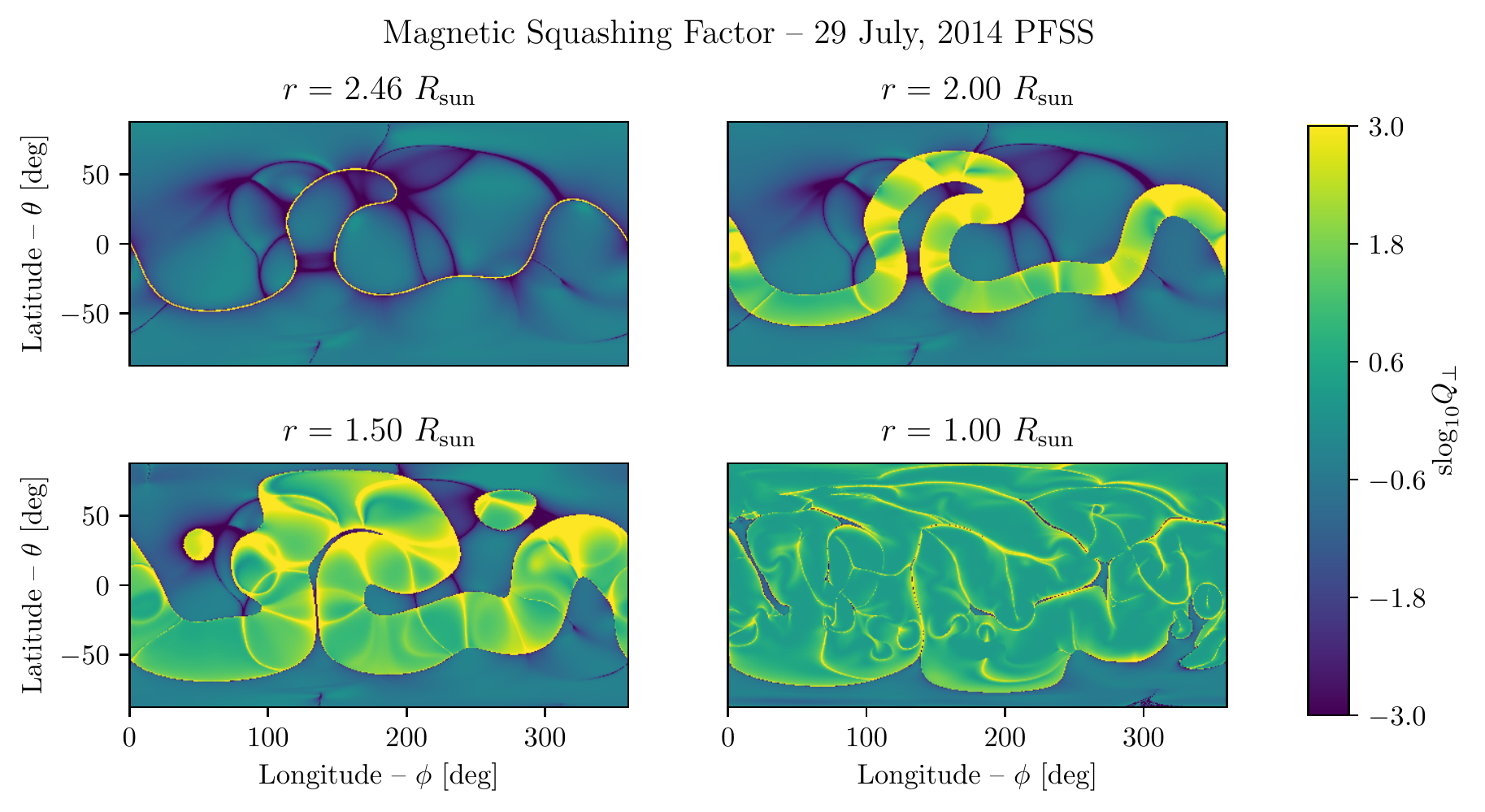}
\caption{Color map of ${\rm slog}_{10} Q_\perp$ at radial positions of $r/R_{\rm sun} = \{1.0, 1.5, 2.0, 2.5 \}$ for the 29 July, 2014 PFSS model. Positive (negative) values indicate closed (open) flux. Dark bands at the upper boundary indicate the intersections of HQVs with the source surface. The presence of separatrix surfaces that divide open and closed flux can be inferred from the abrupt transitions in color and brightness.}
\label{slog10q_multi.fig}
\end{figure*}

For estimating the magnetic squashing factor we use the \texttt{qslSquasher} code \citep{Tassev:2016, Scott:2017a}, which employs GPU computing to do massively parallel field line tracing, as needed for a high-resolution volume rendering. 
Field line integration is performed using trilinear interpolation with an Eulerian integration step size of approximately 0.25 Mm (equivalent to 20 steps per grid cell of the source field). 
The output grid is $120\times 480\times 960$ in Radius ($r$), North Latitude ($\theta$), and Longitude ($\phi$), respectively, with uniform spacing in angular coordinates and exponential spacing in the radial coordinate. 
Since the field-line integrator must be called separately at each grid point, each instance of the global calculation corresponds to $108\times10^6$ individual estimates of $Q_\perp$, each of which involves integration of a 9-dimensional ODE, both forward and backward from the point of interest to the boundary.
Despite the computational advantages of the \texttt{qslSquasher} routine, each global calculation requires several hours to complete on an nVidia Tesla K40 GPU, so this method is useful for studies of up to several tens of individual models, but would be ill-suited to decadal surveys of daily magnetogram models. 

As we have calculated $Q_\perp$ in a volumetric sense, we are able to identify HQVs in three dimensions, and thereby infer the presence of separatrix surfaces and hyperbolic flux tubes based on their morphology. 
We focus primarily on HQVs that intersect the source surface, disregarding some of the more complicated QSL formations associated with twisted and braided flux tubes and sigmoid structures \citep[e.g.,][]{Savcheva:2011gi}. 
Accordingly, where the HQVs have a sheetlike structure, and where $Q_\perp$ is large (but not too large), we expect the underlying field structure to resemble that of an HFT and associated QSL (see Figure 7 of \cite{Titov:2011.731} and Figures 3 and 4 of \cite{Antiochos:2011.732}).
Where $Q_\perp$ is very large (formally infinite), we expect there to be a magnetic null and an associated separatrix surface. 
And, where $Q_\perp$ is discontinuous we expect there to be a bald patch and an associated separatrix surface \citep{Titov:2007gn}. 
Furthermore, the separatrix surface associated with the GHS can be identified as the interface between open and closed flux domains, and while the footpoint locations of individual field lines are not retained by \texttt{qslSquasher}, this information is encoded into the value of $\rm{slog}_{10} Q_\perp = \pm \log_{10} Q_\perp$, which we define to be positive (negative) if the associated field line is closed (open).

Figure \ref{slog10q_multi.fig} shows ${\rm slog}_{10} Q_\perp$ at four different constant-radius slices, from just inside the source surface, down to the photosphere.
The position of the outermost slice is chosen to allow for some closed flux to be visible (thereby indicating the apex of the GHS), which is not possible at the source surface, where all flux is open by construction. 
The color scheme is such that green regions have relatively low $Q_\perp$, while regions of large $Q_\perp$ are either bright yellow or dark violet depending on whether the flux is closed or open. 
In the lower right panel, the dark regions show the photospheric footprint of the various coronal holes, and it is noteworthy that no polar crown is present in the northern hemisphere, owing to the fact that the flux in northern polar region is opposite that of the global dipole field.

\section{Observed S-web Structures}\label{obs_hqa.sec}

We have previously established the convention that regions of sufficiently high $Q_\perp$ ($\gtrsim 10^{3.5}$ in our model) are labeled HQVs, and these may in fact be either QSLs or separatrix surfaces, depending on the presence of magnetic nulls or bald patches. 
The combined network of all HQVs is understood to form the S-web \citep{Antiochos:2011.732}.
At the outermost boundary, where the S-web intersects the source surface, we typically observe elongated, quasi-1D bands of highly squashed flux, which we call ``high-Q arcs'' (HQAs).
HQAs are the intersection of HQVs with the source surface, and these shall be the focus in this investigation, for which the main line of inquiry is the following.
Can we rigorously differentiate between the kinds of structures that form simple arcs and those that form more complicated branching structures, and, if so, what does this tell us about the underlying coronal magnetic field?

The most prominent HQV in the S-web is associated with the HCS, which forms a closed curve that divides positive open flux from negative open flux on the source surface (see the top-left panel of Figure \ref{slog10q_multi.fig} and compare to the top panel of Figure \ref{source_field.fig}). 
The HCS is itself the apex of the GHS, dividing open and closed flux within the coronal volume. 
This is the only HQV that coincides with a polarity inversion line at the outer boundary, and traces out the only location where the closed field reaches the source surface (since the field at the SSPIL is zero, it represents zero net flux, and can neither be identified as open nor closed). 
While the intersection of the GHS HQV is itself an HQA, we shall refer to it exclusively as the HCS to avoid confusion. 

At the source surface all HQAs  -- excluding the HCS -- are necessarily within the open field region, being bounded on either side by open magnetic flux, yet the underlying magnetic field can be highly non-trivial, involving both open and closed flux domains. 
In order to infer the structure of the underlying field, we find it useful to define a categorical scheme, based on how the observed HQAs connect to each other and to the HCS. 
In the following, we shall use the term ``vertex'' to describe the intersection of multiple HQAs, away from the HCS. 
The sections of HQA between intersections are called ``segments'', and these can be bounded on either end by a vertex, or by an intersection with the HCS, or in some cases they may end abruptly, away from any other segment. 
In order that vertices should be defined unambiguously, we require that for segments to join at a vertex there must be at least one sharp angle of intersection -- i.e., a single smooth arc cannot be divided into a pair of segments and a vertex.

\begin{figure*}[ht]
\center
\includegraphics[width=\linewidth]{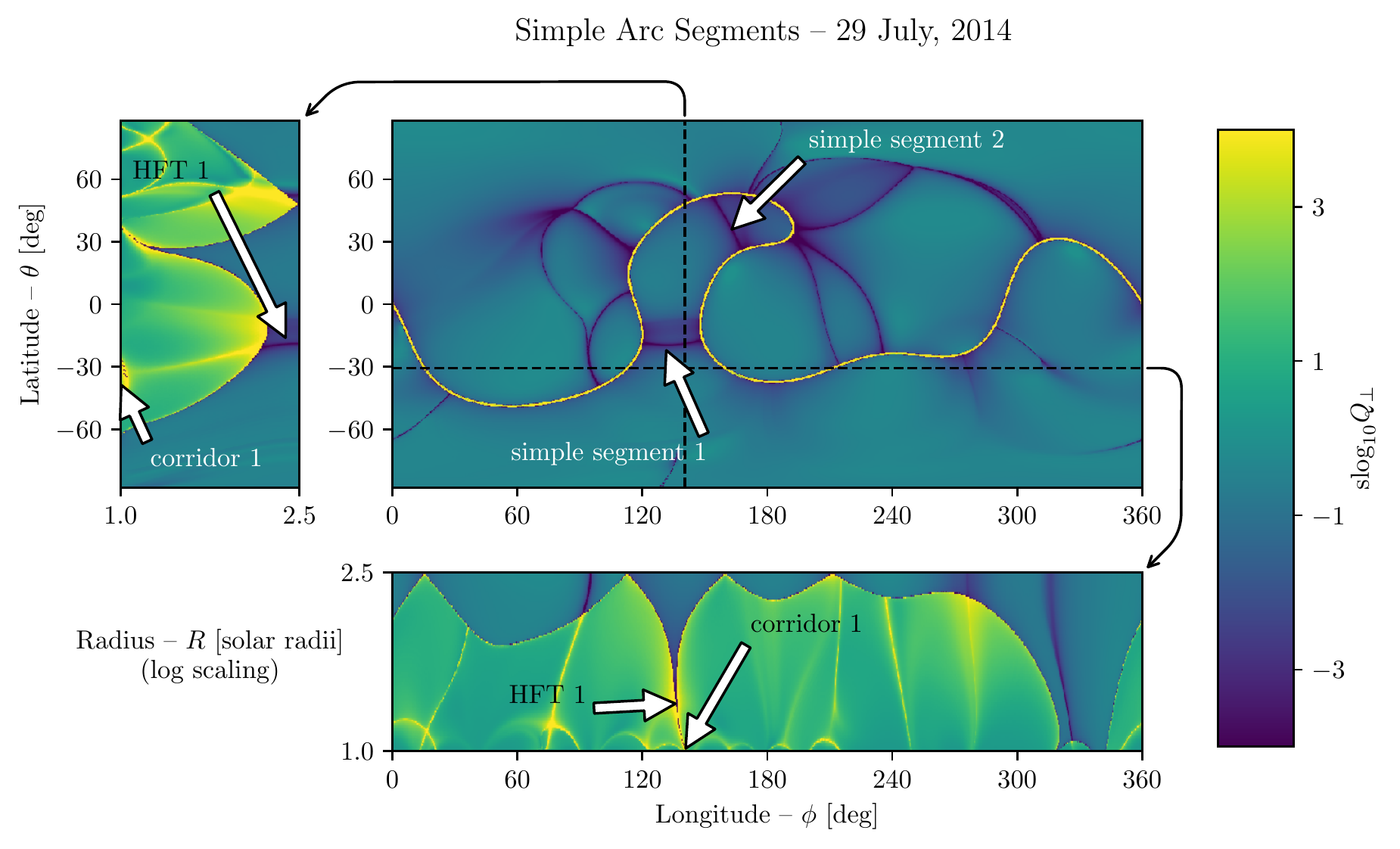}
\caption{Simple arc segments at $r \approx 2.5 R_{\rm sun}$ (source surface). Cross-sections through $\theta \approx -30^\circ$ and $\phi \approx 140^\circ$ show the 3D structure of an HQV that maps down to a narrow corridor, which is bounded by the GHS on either side.}
\label{simp-seg.fig}
\end{figure*}

\subsection{Simple Arc Segments}\label{simp-seg.sec}

First, we consider HQA segments that connect to the HCS at both ends, and have no other intersections, and these we call ``simple segments''. 
Simple HQA segments are ubiquitous in the model, and we have indicated two examples in Figure \ref{simp-seg.fig}. In the figure, the HCS exhibits an obvious excursion to high-latitude at $\phi \approx 130^\circ$. 
If we consider a closed curve comprised of this northerly excursion of the HCS and the HQA labeled ``simple segment 1'', this curve encloses a unipolar negative flux domain that extends into the northern hemisphere.
And within this domain, there is another simple HQA segment, labeled ``simple segment 2'', which encloses a smaller subset of the same flux. 
We refer to the smaller flux domain as being ``embedded'' within the larger, and the associated simple arc segments are said to be ``nested''.

Looking at the bottom and left side panels of Figure \ref{simp-seg.fig}, we see slices through the domain at constant $\theta \approx -30^\circ$ and $\phi \approx 140^\circ$, corresponding to the dashed lines on the main panel. 
From these it is apparent that the HQA labeled ``simple segment 1'' is the footprint of an HQV that extends down through the volume and eventually connects to an open flux corridor on the photosphere. 
Accordingly, we identify this as an HFT, though it may in fact contain nulls at very low heights above the photosphere.
This structure can also be seen by following the HQVs of the two simple segments down through the volume in Figure \ref{slog10q_multi.fig}, where we find that as more of each radial slice is taken up with closed field beneath the GHS, the simple segments contract along their length and expand along their width.
These eventually thin to narrow corridors of open flux at the photosphere, which connect a series of small coronal holes that extend from the southern polar crown into the predominantly closed mid-latitude region. 

In contrast, we now consider segments that do not connect twice the to SSPIL, and these we divide into two classes: detached, and branching.

\begin{figure*}[ht]
\center
\includegraphics[width=\linewidth]{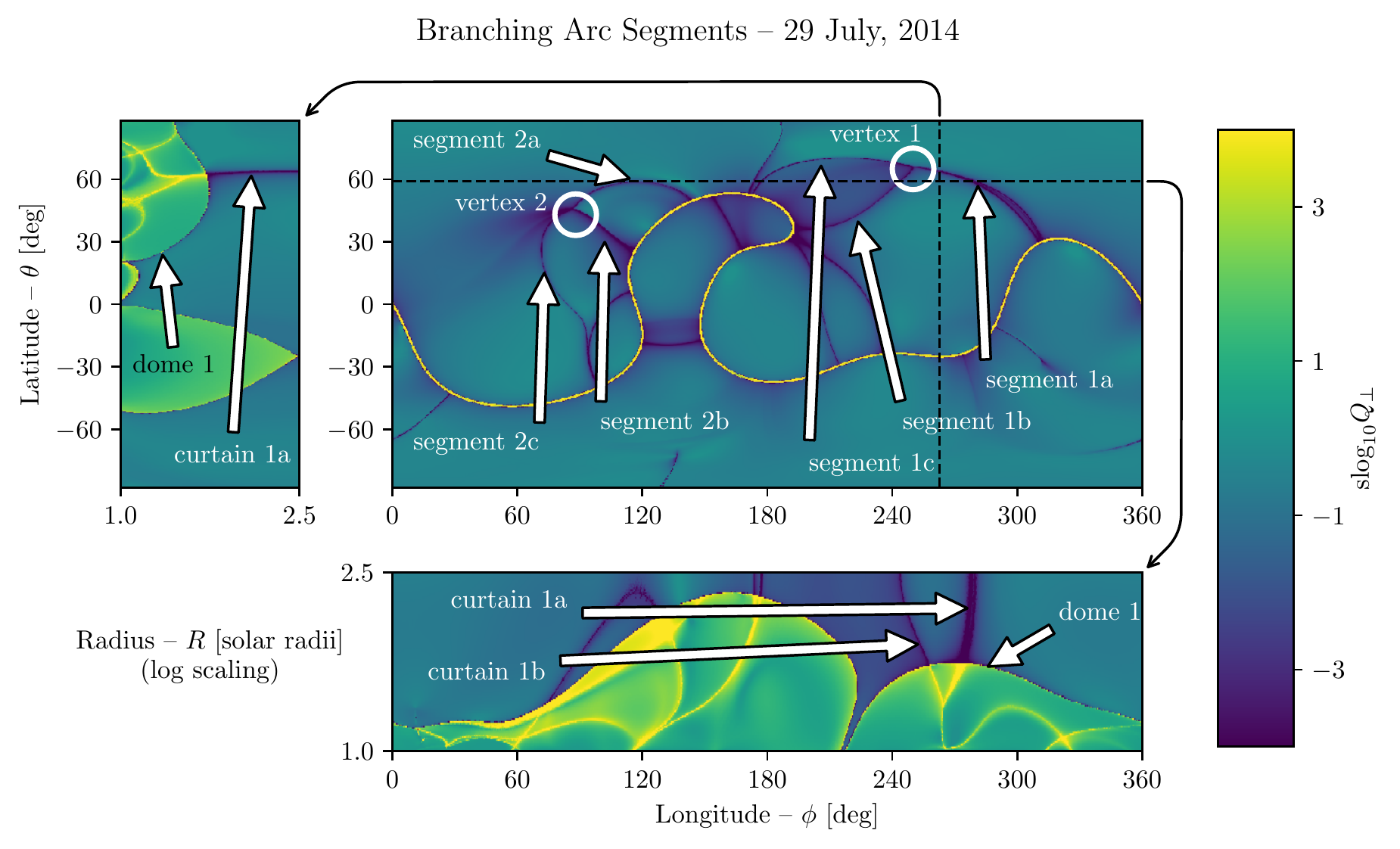}
\caption{A group of branching segments at the $r=2.5R_{\rm sun}$ surface (source surface). Cross sections through $\theta \approx 60^\circ$ and $\phi \approx 260^\circ$ show the 3D structure of an HQV that maps to a separatrix dome whose apex is at $r\approx1.75R_{\rm sun}$.}
\label{brch-seg.fig}
\end{figure*}

\subsection{Branching Arc Segments}\label{brch-seg.sec}

By contrast with simple segments, which have exactly two points of intersection with the HCS and no other intersections, we refer to ``branching segments'' as HQA segments that intersect each other at a vertex, away from the HCS. 
The most common arrangement of branching segments is a triplet, in which each segment has a single intersection with the HCS, while the other end terminates in a vertex that is shared by all three. 
This is exemplified in Figure \ref{brch-seg.fig}, which shows two vertices, labeled ``vertex 1'' and ``vertex 2'', each of which is comprised of three branching segments, labeled ``segment 1a'', ``segment 1b'', etc.
In the case of ``vertex 1'', the segment labeled ``1a'' appears to be formed of two merged segments, so this branch system may involve four segments, rather than three.
In the case of ``vertex 2'', the segment labeled ``2c'' appears to merge with a simple segment farther to the south. 
Variations of this kind from one branching system to the next are common, but do not appear to materially affect the structure of the field near the vertices.

We first note from Figure \ref{brch-seg.fig} that, where simple segments have been observed to occur on the concave side of HCS excursions, these branching segments tend to occur on the convex side of the HCS, and this has been found to be generally true for the other models that we have considered.
Furthermore, on inspecting the bottom and left panels in Figure \ref{brch-seg.fig}, it is apparent that, whereas the simple segments tend to be associated with HFTs and narrow corridors, the HQVs associated with branching segments tend to map down to closed separatrix domes, leading us to infer that these may support large scale topological structures, rather than simple HFTs. 
In the figure, the HQVs that connect to the vertices have been labeled as ``curtains'', owing to their morphological similarity to the separatrix fan curtains described in \cite{Titov:2011.731}, and especially those depicted in Figures 3, 4, and 5 of \cite{Platten:2014j}.
The association between vertices and separatrix dome structures can also be seen in Figure \ref{slog10q_multi.fig}, in which both of the mentioned vertices can be traced down to small circular patches of closed flux in the bottom-left panel, these being the apexes of closed field separatrix domes.

\begin{figure*}[ht]
\center
\includegraphics[width=\linewidth]{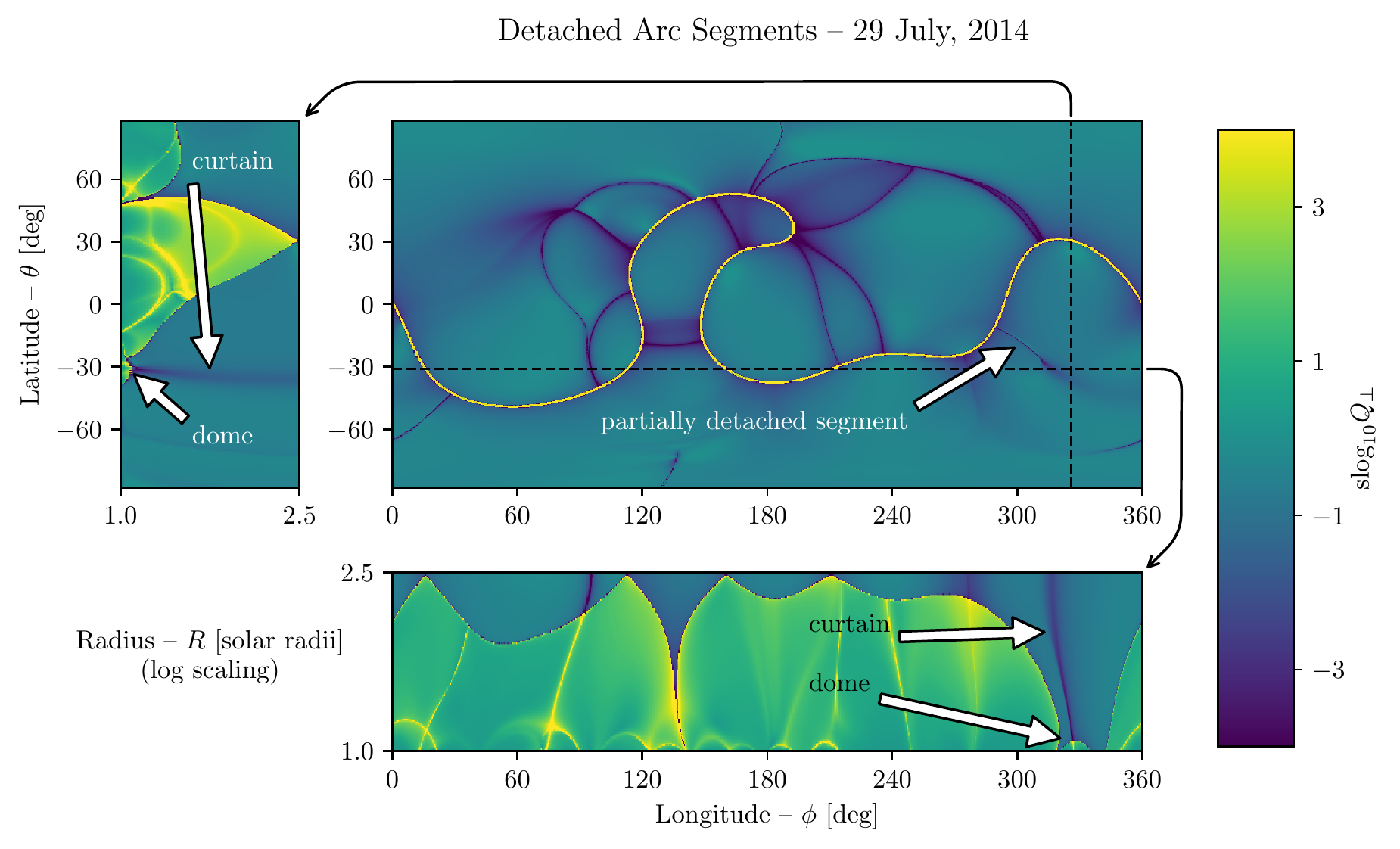}
\caption{A partially detached segment at the $r=2.5R_{\rm sun}$ surface (source surface). The segment forms has only one intersection, and terminates away from the HCS. Cross sections through $\theta \approx -30^\circ$ and $\phi \approx 320^\circ$ show the 3D structure of an HQV that maps to a separatrix dome whose apex is at $r\approx1.1R_{\rm sun}$.}
\label{dtch-seg.fig}
\end{figure*}

\subsection{Detached Arc Segments}\label{dtch-seg.sec}

Lastly, we address detached segments, which are HQAs with at least one free end, which terminates away from any other S-web structure. 
These are referred to as ``partially detached'' in cases where the segment has one intersection and one free end, or ``fully detached'' in cases where the segment has two free ends and no intersections. 
In principle, a partially detached segment could join a branching group at a vertex; however, this has not been observed.
Detached segments appear far less commonly than do simple or branching segments; nonetheless, these structures are generic features, and, as we shall see, they are critical to our overall understanding of HQA formation. 

There are three examples of detached segments in our global model. 
One of these is in the northern hemisphere, at $\phi \approx 180^\circ$, and extends from the HCS to the northern pole.
Another is in the southern hemisphere, at $\phi \approx 130^\circ$, and exhibits a similar polar connection but forms no connection with the HCS.
Both of these are considered unreliable, as they involve field line integration through the polar region, where the model magnetic field is poorly constrained.

The third example is indicated in Figure \ref{dtch-seg.fig}, and exhibits one connection to the HCS, extending south-west from there and terminating abruptly, with no other intersections. 
Like the branching segments previously discussed, this detached segment also forms a ``curtain'', which is morphologically similar to the configuration in Figure 4 of \cite{Platten:2014j}, intersecting the GHS at one end and terminating above the apex of a closed field separatrix dome in the low corona. 
Examples from other models show a similar tendancy, with the termination point typically mapping down onto a closed field separatrix dome, which is usually smaller than for branching segments, and typically occurs at lower heights in the corona. 
As there are so few examples of detached segments, it is also difficult to say whether these form preferentially on the concave or convex side of the HCS, although we can say that no nested examples have been observed.

\subsection{Summary of Observed Structures}

In summary, we have found that the manner in which HQA segments intersect the HCS and each other can be an indicator of the structure of the interior of the domain. 
In particular, where the HCS has a large excursion away from mid-latitude, the concave side of the HCS, which corresponds to the intrusion of open flux into the opposite-signed hemisphere, typically supports simple arc segments, which likely correspond to hyperbolic flux tubes that map to narrow corridors of open flux at the photosphere. Conversely, the convex side of the HCS, which corresponds to the flux that has been deformed away from mid-latitude to accommodate the deformation of the GHS, typically supports branching arc segments, which intersect at vertices, away from the HCS, and typically connect to a closed-flux dome structures lower down in the coronal volume.

\section{Theoretical Interpretation}\label{emp_hqa.sec}

While the associations made in the previous section are plausible, the analysis lacks rigour -- we have drawn heavily on visual comparisons to theoretical studies of separatrix surface topologies and analytical field models, but we have not constructed a magnetic skeleton, so our only well defined topological feature is the open closed boundary. 
We now focus on the formation of such structures from a theoretical perspective, by considering the field lines that form the OCB, and how these map between the photosphere and the source-surface.

\subsection{Narrow Corridors}

\begin{figure}[ht]
\center
\includegraphics[width=\linewidth]{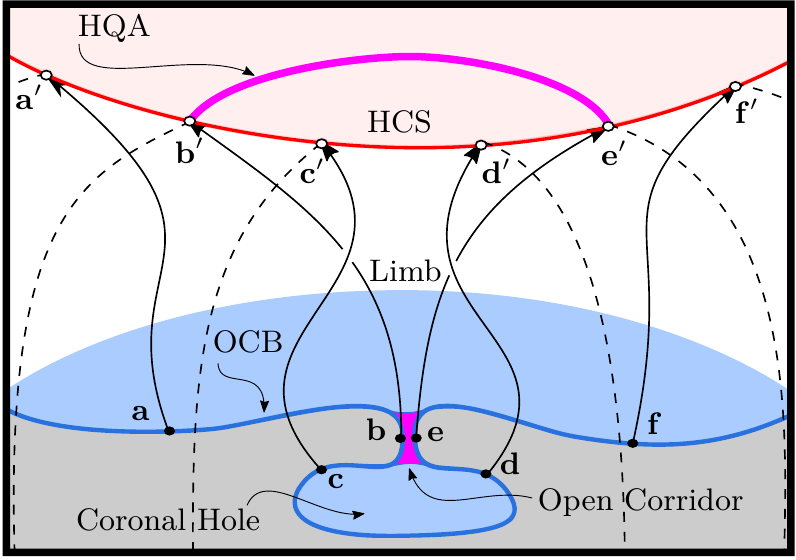}
\caption{A QSL is formed in a narrow corridor where two sections of the open-closed boundary come to close proximity. Because the OCB is straightened and elongated on the source surface, flux within the narrow corridor forms a hyperbolic flux-tube whose imprint on the source surface is an HQA.}
\label{simp-corr.fig}
\end{figure}

In order to inform our understanding of HQAs we first consider the simplest configuration that can create an HQV with an imprint at the source surface, that being a QSL formed from an HFT along the OCB. 
Such a configuration was described by \cite{Antiochos:2011.732}, and consists of a single meandering OCB line that passes near to itself, creating an isolated coronal hole which remains connected to the larger region of open flux through a narrow corridor, bounded on either side by the closed flux beneath the GHS. 
A depiction of this configuration is shown in Figure \ref{simp-corr.fig} \citep[see also Figure 1 of ][]{Antiochos:2011.732}. 
We assume that apart from this feature the magnetic field is the simple global dipole field. In this configuration, the entire OCB at the photosphere is a pair of simple closed curves, one in each magnetic polarity (being the photospheric footprint of the GHS separatrix surface). 
Field lines map points on each of these curves to the HCS in a one-to-one fashion that preserves the ordering of field line footpoints (mathematically the HCS is diffemorphic to each of the OCB footprints), this being critical for the following. 

Now consider the mapping between the OCB footprint shown in Figure \ref{simp-corr.fig} (marked in blue) and the HCS (red curve). 
As shown, points ${\bf b}$ and ${\bf e}$ on either side of the narrow corridor (magenta) must map to points ${\bf b}^\prime$ and ${\bf e}^\prime$ far separated on the SSPIL to preserve the ordering of the mapping. 
In addition, the flux from within the narrow corridor must connect between ${\bf b}^\prime$ and ${\bf e}^\prime$ at the source surface, and moreover must enclose all of the flux from the isolated coronal hole. 
This means that this corridor flux must be stretched out to form the magenta HQA shown in the figure at the top boundary, creating an HFT.

By construction, an HFT formed from a single open flux corridor must create an HQA that intersects the HCS at both ends. We submit that, by extension, any configuration of linked corridors and their corresponding HFTs will create a network of multiple HQAs, all of which connect twice to the HCS. 
These may merge, and share common points of intersection with the HCS, or may nest, so that larger coronal holes contain HQAs that subdivide the region into smaller coronal holes, but the assertion is that no vertices will be formed away from the HCS. 

\begin{figure}[ht]
\center
\includegraphics[width=\linewidth]{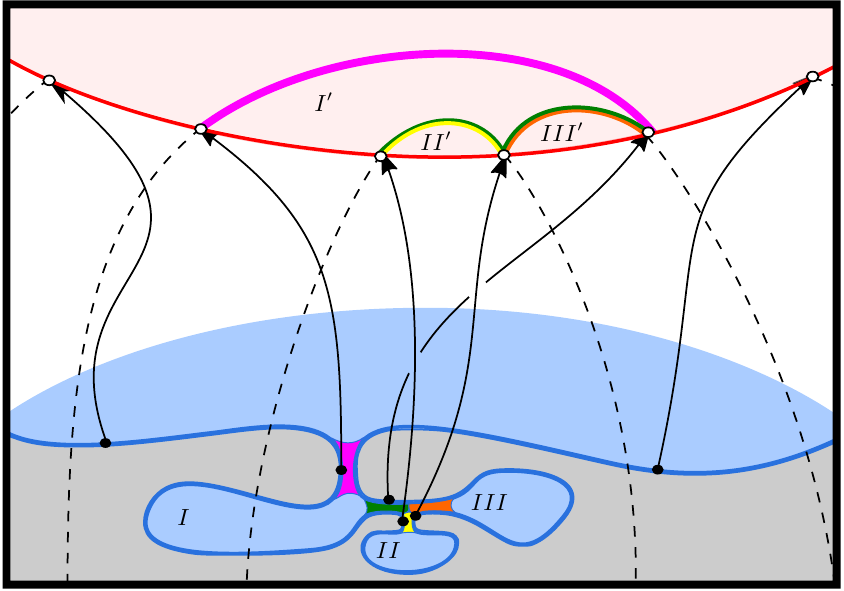}
\caption{Multiple Coronal holes linked in a system of narrow corridors. Depending on the configuration, these can result in HQAs that are nested or adjacent, with distinct or shared intersections with the HCS. In all cases, the HQAs form simple arcs segments.}
\label{mult-corr.fig}
\end{figure}

To support this claim, we have illustrated a complex corridor system in Figure \ref{mult-corr.fig}, which contains a generic combination of HQAs, resulting from corridors that link coronal hole domains in parallel or sequential order. 
As in Figure \ref{simp-corr.fig}, the OCB (in the polarity shown) is a single, continuous curve, and it must map to the HCS in a way that preserves the ordering of field-line footpoints. 
Additionally, the open photospheric flux domains labeled $I$, $II$, and $III$  map to domains $I^\prime$, $II^\prime$, and $III^\prime$ at the source surface. 

Since points along either side of the magenta corridor are the limits of a line segment containing points along either side of the green, yellow, and orange corridors, the corresponding HQAs from these latter three must be nested within the former. 
Coronal holes $II$ and $III$ are connected in parallel (via the yellow and orange corridors), and together are connected in series (via the green corridor) with coronal hole $I$. 
Thus, the open flux domains $II^\prime$ and $III^\prime$ are adjacent to each other and are embedded within $I^\prime$.

Because the green corridor is adjacent to the magenta at one end, the corresponding HQAs appear to merge at one end, forming a common intersection with the HCS. 
The yellow and orange corridors are similarly adjacent at the photosphere, and their HQAs share a common point of intersection with the HCS as well, though they are adjacent rather than nested. 
More subtle is the HQA corresponding to the green corridor, which partitions no flux in itself, so there can be no flux between the green HQA and the orange and yellow HQAs. 
The green corridor is effectively a shared ``upstream'' corridor region, which maps to a double-humped structure that hugs the top of the orange and yellow HQAs, thickening each but not changing their general shape. 

The construction in Figure \ref{mult-corr.fig} is intended to be an extreme example, which is more complex than is typically observed in the coronal magnetic field extrapolations that we have considered. 
Nonetheless, this example serves to illustrate that in any configuration in which the photospheric OCB forms a simple closed curve -- irrespective of its geometrical complexity -- the resulting HFTs cannot form HQA vertices away from the HCS.

\subsection{Null Dome Configurations}\label{null_top.subsec}

To understand how vertices of HQAs are formed, we must consider structures that map segments of the OCB into unipolar, open flux domains, away from the HCS, which necessitates that there should be some singular behavior in the mapping, and this leads to the consideration of magnetic nulls. 
A survey of magnetic null topologies in global models was previously performed by \cite{Platten:2014j}, who considered various configurations of coronal nulls and separatrix surfaces. 
The simplest of these structures is a so-called separatrix dome, which is a closed separatrix surface that is formed when a coronal null exhibits a fan (separatrix) surface that curves down on all sides and intersects the photosphere along a simple closed curve \citep[see Fig. 2 of ][]{Platten:2014j}.
Underneath, and completely enclosed by the dome, there must exist a patch of parasitic polarity, which closes down locally, necessitating a separatrix surface to divide it from the surrounding flux, which maps to a more distant region.
Reconnection at these generic structures in general permits an exchange of plasma/magnetic flux between the inside and outside of the dome \citep{Pontin:2007a,Pontin:2013}, and has been invoked to explain a range of energetic coronal events. 

\begin{figure}[ht]
\center
\includegraphics[width=\linewidth]{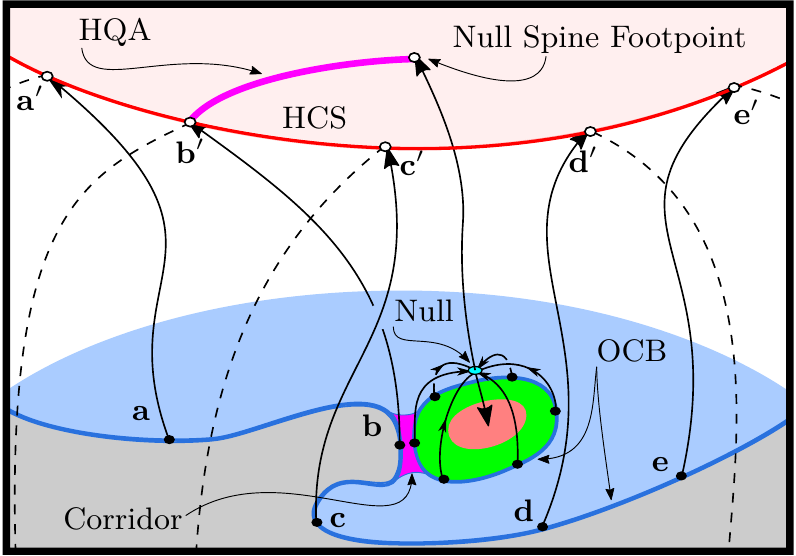}
\caption{A QSL formed in a narrow corridor where the main OCB is brought close to a null separatrix dome footprint, formed above a parasitic polarity region (salmon oval) in an otherwise open coronal hole. An HFT is formed, which extends from the corridor to an HQA at the source surface; however, because the null spine is not positioned along the HCS, the HQA terminates abruptly in the open field region.}
\label{simp-spine.fig}
\end{figure}

The field line mapping in a magnetic dome topology with a single null is such that any point along the footprint of the dome maps to the null point, and from there to the footpoints of the null spine line: the `inner spine' footpoint lies in the parasitic polarity, while the `outer spine' foopoint can be either open or closed depending on the nature of the surrounding flux.
If the outer spine line is open then the dome separates open flux from closed, so its footprint is a part of the OCB, but is disconnected from the primary OCB that forms along the footprint of the GHS (see Figure \ref{simp-spine.fig}). The entirety of this detached segment of the OCB maps to the null spine footpoint, which terminates somewhere in the open flux region, away from the HCS. 
It follows, therefore, that if an HFT is formed adjacent to a separatrix dome with an open spine line (say, due to the presence of a narrow open field corridor between the null dome and the GHS) then one end of the corresponding HQA will terminate at the null spine line footpoint, away from the HCS, as shown in Figure \ref{simp-spine.fig}.

This consideration for the portion of the OCB that corresponds to the footprint of null fan surfaces is critical to our understanding of HQAs. 
Because the fan surface footprint forms a non-zero portion of the OCB that is mapped to a point \emph{away from} the HCS, these structures allow narrow corridors of open flux to be bounded on both sides by regions of closed field, without requiring that the corresponding HQA have two intersections with the HCS. 
The configuration described above and depicted in Figure \ref{simp-spine.fig} provides a plausible explanation for the detached segments described in section \ref{dtch-seg.sec} and depicted in Figure \ref{dtch-seg.fig}, and this provides a key element for our understanding of more complicated structures.

The inclusion of nulls in the model is not difficult to motivate, as it has been shown that coronal nulls are generic and ubiquitous, especially in the low corona \citep{Longcope:2009.soph, Cook:2009in, Freed:2015, Edwards:2015b}; however, the connection to narrow corridors is not immediately obvious until one considers why narrow corridors form in the first place. 
Referring to the bottom-left panel of Figure \ref{slog10q_multi.fig}, we see that adjacent to the apex of the closed dome structure at $(\theta, \phi) \approx (30^\circ, 50^\circ)$, there is an ellipsoidal dome structure (closed curve of high Q) within the GHS and centered at $(\theta, \phi) \approx (10^\circ, 90^\circ)$, which causes the GHS to bulge into the open field region. 
As it happens, this is a rather generic feature, with corridors commonly forming in regions where separatrix dome structures in adjacent closed field domains press toward one another, pinching the open flux between them.
The single null dome topology described above is only the simplest possible configuration with magnetic nulls on the OCB. 
Much more complex structures are possible in which closed dome structures comprised of the fan surfaces of multiple nulls are present \citep{Titov:2011.731,Platten:2014j}. 

\begin{figure*}
\center
\includegraphics[width=0.75\linewidth]{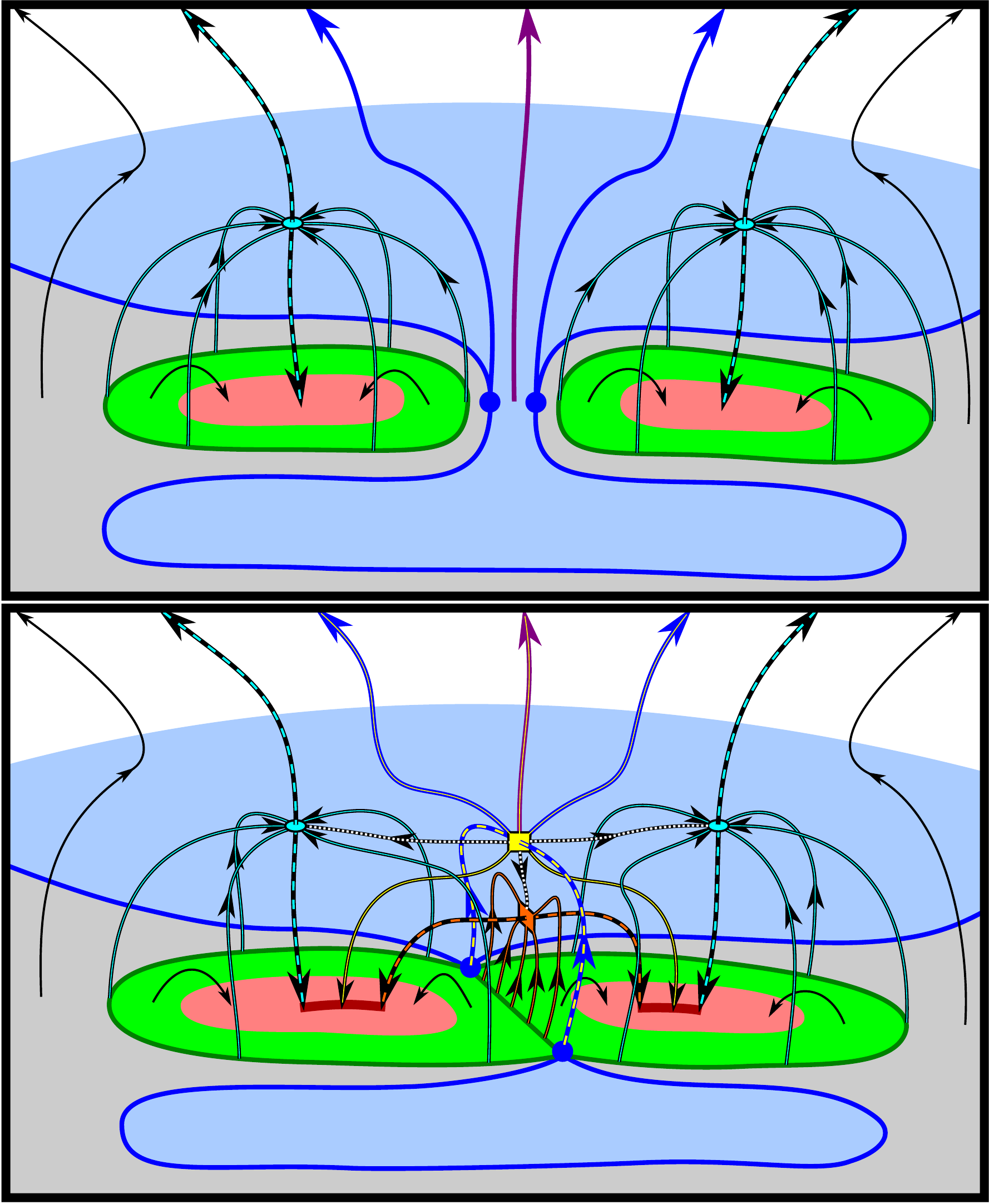}
\caption{Diagram of a detached coronal hole. The OCB (bold blue curve) is the footprint of the GHS, which maps down from the HCS (not shown) along the bold blue field lines. Black field lines represent closed flux. Field lines with color overlay map to magnetic nulls of the same color coding, with dashed overlay indicating spine lines and solid overlay indicating fan field lines. Positive open flux is depicted in light blue, while the bright green and salmon colored regions indicate positive and negative flux within the null separatrix domes. In the top panel there are only two nulls, each within the closed field region and the smaller coronal hole remains connected to the larger via an open flux corridor. In the bottom panel, the open flux of the corridor is replaced by a pair of nulls and their fan surface field lines. The coronal holes are now ``linked but not connected''.}
\label{corridor-collapse.fig}
\end{figure*}

We have explored the interaction of a pair of separatrix domes in the closed field, which form a corridor that is made progressively narrower, to the point of having zero width. 
This was done using an analytical potential field model involving a pair of submerged monopoles and a uniform background field. 
Our findings are illustrated in Figure \ref{corridor-collapse.fig} in which we show two configurations with nearly identical flux distributions that both create a small coronal hole, which is separated from the majority open flux domain by either a narrow corridor or a separatrix surface, depending on the details of the local magnetic field.

In the top panel of Figure \ref{corridor-collapse.fig} the purple field line in the center represents open flux within a narrow corridor, which is bounded on either side by the GHS separatrix surface.
There are two nulls in the closed field region, and none in the open field region, so the topology is relatively simple, and for sufficient narrowing of the corridor the imprint on the source surface would be that of a simple HQA, similar to Figure \ref{simp-corr.fig}. 
If, however, the corridor is narrowed to zero width, either due to motion of the parasitic polarity regions (salmon) or due to changes in the large-scale field, a pair of nulls (of opposite type and zero combined topological degree) can form within the corridor via a saddle-node bifurcation \citep{Priest:1996d}.
This leads to a configuration like the one shown in the bottom panel of Figure \ref{corridor-collapse.fig}, in which the HFT of the previous example has been replaced by a pair of intersecting separatrix fan surfaces from the two central nulls.

The fan surface of the upper central null (yellow rectangle) maps up to the HQA formerly associated with the corridor HFT and intersects the GHS along the bold blue field lines.
It is bounded by the spine lines of the pre-existing (cyan) nulls, making it a ``curtain'', which spans from the closed field, across the open field, and back into the closed field again.
The fan surfaces of the cyan nulls intersect the fan surface of the yellow null along separator field lines (dashed white), and together these surfaces form a single separatrix dome, being comprised of the two pre-existing cyan null domes, which are joined along the spines of the yellow null.
The presence of the lower central null is required by the separation of the two parasitic polarities, and if these are merged (say due to flux cacellation), the orange null submerges, leaving a simpler triple-null system.\footnote{Other variations that arise from similar source configurations include the merging of the upper nulls via the reverse of either a pitchfork or saddle-node bifurcation, but these are less relevant to the discussion of HQAs.}

Because the spine lines of the cyan nulls are closed, the triple-null dome is entirely in the closed field, \emph{except} along the spines of the (yellow) central null, which form an ``archway'' that ``bridges'' the region previously associated with the narrow corridor.
These spines (yellow dashed) form a portion of the GHS and thus map to a pair of points (blue dots) on the OCB, which are the only points of intersection between the OCB and the null dome footprint. 
The coronal hole that is formed in this case is, therefore, now connected to the majority flux domain only through the spine lines of the upper central null, making it ``linked but not connected'', much like the configuration described in \cite{Titov:2011.731}.

All of these states are dynamically connected through a sequence of potential fields, however the details are strongly dependent on the model parameters, especially the symmetry of the source configuration. 
Moreover, the construction that we have described, arising from the spontaneous formation of nulls, is only one way of forming a triple-null dome whose fan spans the OCB.
Another way would be to begin with a triple null system where both vertical spines are embedded in the same closed field region, and then to let one spine line open across the GHS surface, migrate across the corridor, and then close down across the opposing GHS surface \citep[c.f.][]{Titov:2011.731}.  
It is enough to know that both configurations depicted in Figure \ref{corridor-collapse.fig} are generic and are likely present in the corona on some scale at all times.

\begin{figure}
\center
\includegraphics[width=\linewidth]{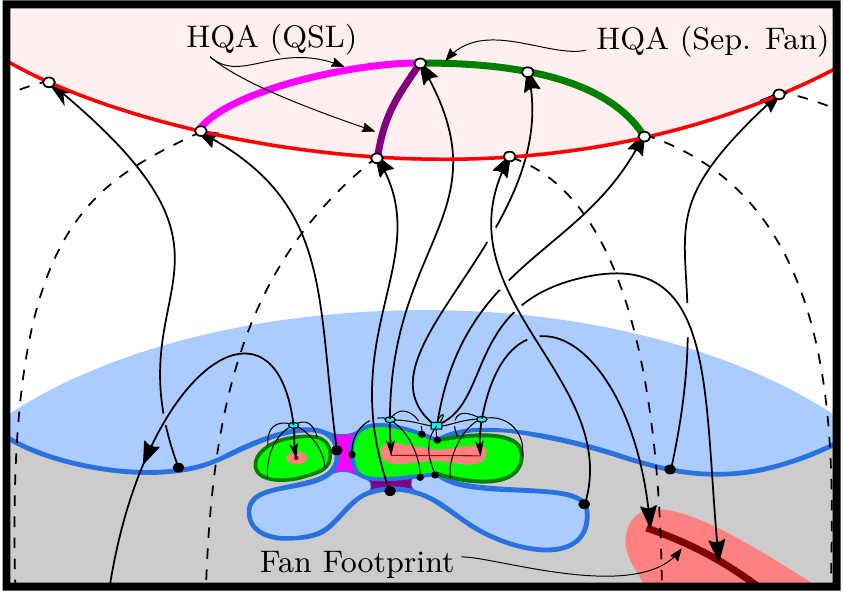}
\caption{A system of HQAs formed from a combination of HFTs and separatrix surfaces, with a vertex at the spine line of a magnetic null in the open field. Three HQAs meet at the vertex, two being formed from HFTs, and a third being the footprint of a separatrix ``curtain'' from a triple null system that spans the OCB. }
\label{mult-spine.fig}
\end{figure}

Allowing for the inclusion of nulls with spine lines in the open field, as well as linked coronal holes separated by multi-null separatrix domes, we can now consider a generic configuration with multiple arcs meeting at a vertex away from the HCS. 
In Figure \ref{simp-spine.fig} we showed how a single arc could terminate away from the HCS, but given that a single separatrix dome footprint can form multiple corridors with different portions of the OCB, it is clear that multiple HQAs can all terminate at a single spine line, creating a vertex as described in section \ref{brch-seg.sec}. 
And by considering the merging of null domes across an open corridor, we have seen these can vanish to zero width, ultimately becoming separatrix surfaces, which leads us to the conclude that the distinction between QSLs and separatrix surfaces in the S-web is largely a matter limiting cases.

With these considerations in mind, we can now consider a generic configuration involving an ensemble of the various structures that we have described, as shown in Figure \ref{mult-spine.fig}. 
In the figure, the OCB is again a single line; however, the triple null system (which has one open outer spine and one closed) has created a `singular segment' along the footprint of the GHS, i.e., a segment that maps to the single footpoint of the open field null spine line, away from the HCS. 
The two narrow corridors, in magenta and purple, form an HQA vertex at the null spine line, where a third HQA segment, formed from the fan curtain of the central null, completes the triple arc system. 

This is only a representative example -- either of the two corridors could also pinch off to create a fully detached (but linked) coronal hole.  
We note that the formation of this `singular segment' of the OCB attached to the main GHS OCB is critically dependent on the presence of a multi-null dome structure that spans the OCB, with at least one closed spine and one open spine: since the intersection of a spine with a separatrix is topologically unstable, any single-null dome must lie either entirely in the open field or entirely in the closed field, away from the GHS OCB \citep{Edmondson:2010.714}. 
We also note that, while the triple null dome contributes to the complexity of this example (allowing one of the HQAs to be a genuine separatrix footprint), it is the presence of the open field spine line, and not the complexity of dome to which it attaches, that allows for the formation of a vertex away from the HCS.

\subsection{Observed S-web Structures: revisited}

\begin{figure*}
\center
\includegraphics[width=\linewidth]{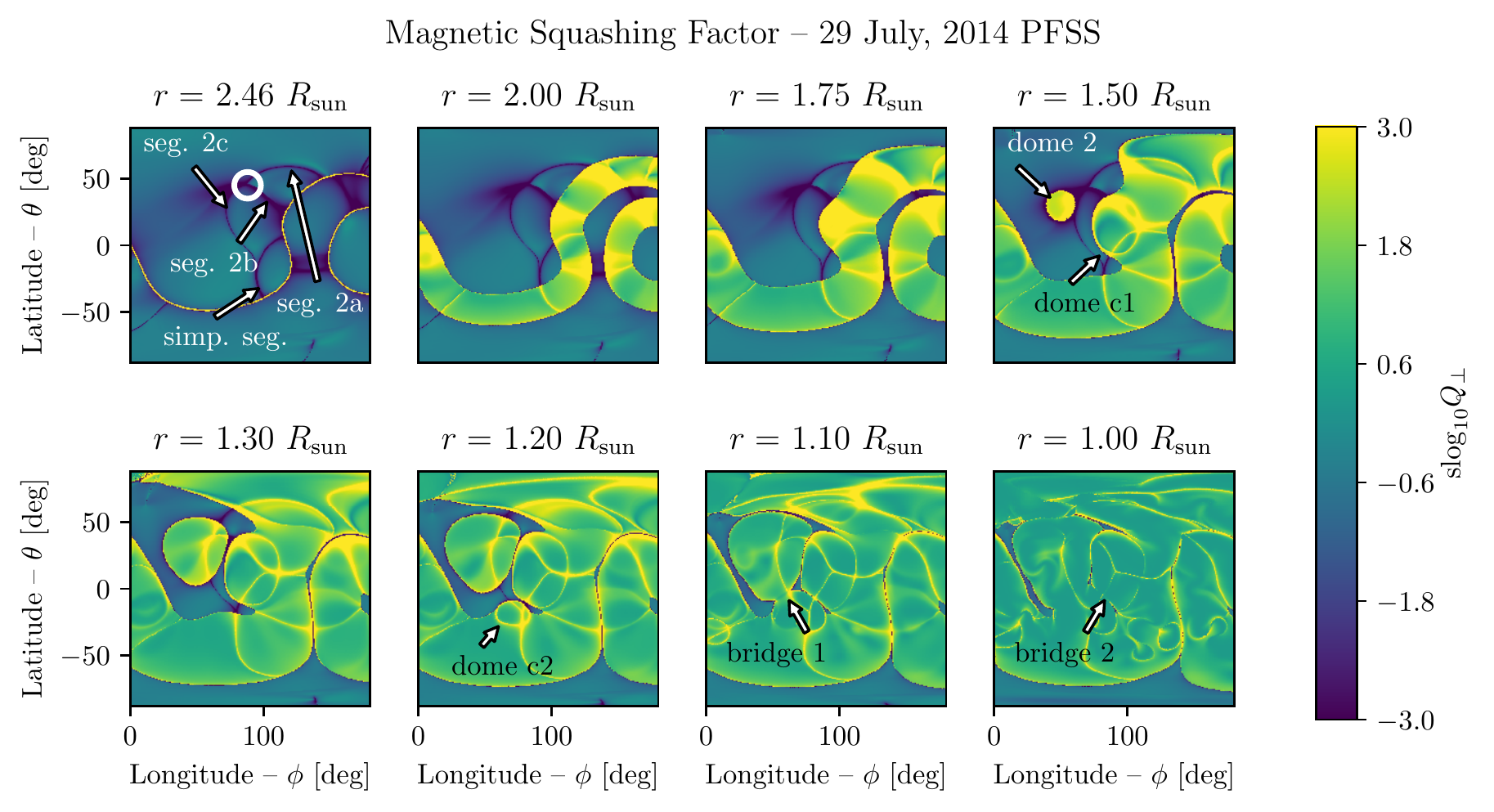}
\caption{Perpendicular squashing factor (${\rm slog}_{10}Q_\perp$) shown in 8 slices of constant radius from photosphere to source surface. The domain is truncated to $\phi \in \{ 0^\circ, 180^\circ \}$ to focus on the system of branching segments with their vertex at $\theta \approx40^\circ$, $\phi \approx 90^\circ$. }
\label{slog10q_mult-left.fig}
\end{figure*}

Following our discussion of HQA vertices as spine-line indicators, we can now return to the global model and interpret the structures that we see there. 
In particular, if we consider the vertex labeled ``vertex 2'' in Figure \ref{brch-seg.fig}, we can now describe the formation of each of the three attached segments (2a, 2b, 2c in the figure) as well as the simple segment that appears to merge with segment 2c near $\theta \approx -20^\circ$.

In Figure \ref{slog10q_mult-left.fig}, the western hemisphere is shown at eight different radial slices. 
The same vertex is indicated by the white circle at $\theta \approx 40^\circ$ and $\phi \approx 90^\circ$ in the top left panel, and this is indicative of a null spine line associated with the closed field separatrix surfaced labeled ``dome 2'' at $\theta \approx 30^\circ$ and $\phi \approx 50^\circ$ in the top-right panel ($r = 1.5 R_{\rm sun}$). 
Another separatrix dome, labeled ``dome c1'' can be seen in the closed field, within the GHS, and this causes the OCB to bulge toward the open-field dome. 

In the lower left two panels ($r = 1. 3 R_{\rm sun}$ and $r = 1.2 R_{\rm sun}$) the closed field region has expanded significantly, and the lower two branching segments (2b and 2c), as well as the simple segment, can each be associated with closed field separatrix domes (c1 and c2). 
The simple segment now clearly spans a corridor bounded by two dome structures within the GHS, while 2b and 2c both span from the isolated dome across to the GHS.
In the $r = 1.1 R_{\rm sun}$ panel (bottom, second from right), the domes connected by segment 2c have now merged (through the region labeled ``bridge 1''), much like the triple null configuration in Figure \ref{mult-spine.fig}, and we now identify this segment as the footprint of a genuine separatrix curtain, which connects to a null dome comprised of (at least) three nulls, the spine from one of which is responsible for the formation of vertex 2. 
In the bottom-right panel of Figure \ref{slog10q_mult-left.fig} we now see that the coronal holes enclosed by segments 2a and 2b remain connected to the larger open flux domain (although their width cannot be accurately resolved), so the example shown in Figure \ref{mult-spine.fig} would seem to be well representative of the system of branching segments that we have described.

We also note that the narrow corridor associated with the simple segment previously mentioned has vanished (at the location labeled ``bridge 2'') between $r = 1.1 R_{\rm sun}$ and $r = 1.0 R_{\rm sun}$, and the two distinct dome structures have seemingly been replaced by a structure similar to that described in Figure \ref{corridor-collapse.fig}.
The merging of separatrix domes at low coronal heights appears to be a generic feature; we find from inspection that the majority of seemingly distinct dome structures are in fact connected by an archway (at near photospheric heights) to one or more adjacent domes.
This increase in complexity near the photosphere is to be expected -- as we have discussed, the introduction of new nulls can cause separatrix surfaces to become linked, and their respective flux domains to merge.
And it has been shown that increased resolution in global models leads to an increasing number of nulls, particularly at radial heights comparable or less than the length scale of the photospheric magnetic field \citep{Longcope:2008db, Edwards:2015b}.
Nonetheless, the importance of the large scale null topology within the GHS should not be underestimated, as it could have a significant effect on rates of reconnection at the open closed boundary, thereby influencing interchange reconnection processes.

\section{Discussion}\label{disc.sec}

The aim of this investigation has been to better understand the origin of high-Q arcs (the `S-web') in global coronal field extrapolations, through comparison with generic model configurations consisting of narrow corridors and magnetic null topologies. 
To this end, we have shown, using the popular PFSS coronal field model, that simple arcs that meet with the heliospheric current sheet at both ends tend to be good indicators of narrow flux corridors as per \cite{Antiochos:2011.732}. 
By contrast, branching structures with vertices that occur away from the heliospheric current sheet require a null spine line in the open field, and are therefore good indicators of more complex null dome topologies, such as those discussed in \cite{Platten:2014j}. 
And we have seen that the distinction between these two regimes is largely a matter of where the nulls are positioned, as even the simpler corridor picture is likely to involve null dome structures in the closed field region. 

The implications of a study such as this are pertinent to future efforts at understanding interchange reconnection and its role in the acceleration/structuring of the slow solar wind. 
Where interchange reconnection involves flux that is bounded only by the global helmet streamer, then the exchange is between open field plasma and plasma that occupies a large coronal volume, out to approximately the Alfv\'en surface. 
However, if volumes associated with null dome topologies that occur beneath low-lying magnetic nulls are seen to participate, then the exchange is likely to occur between open flux and dense plasma from the low corona, such as near active regions. 
This distinction can be important both for the details of the reconnection process, and for the composition of the material that is exchanged. 
We hypothesise therefore that high-Q arc vertices that occur away from the heliospheric current sheet may be preferential locations to find plasma of a low-coronal composition, these being associated with such dome topologies.

In future work, we intend to extend this study through the consideration of the full, 3D structure of the S-web in global models, as well as the extension of our survey to a greater number of model cases, for both potential and, more generically, non-linear force-free fields \citep[for which the underlying magnetic topology is expected to be significantly more complex, as discussed by][]{Edwards:2015a}. 
To this end, we have are developing a segmentation technique, whereby the squashing factor is used to divide the coronal volume into flux domains, with high-Q volumes (being a combination of QSLs and separatrix surfaces) then recovered as the interfaces between these domains. 
In this way, arc vertices, which are the imprints of intersecting separatrix surfaces and hyperbolic flux tubes, can be extracted automatically, and the corresponding volumes can be inspected for the existence of magnetic nulls and other topological features.

\section{Acknowledgements}\label{acknowledgements.sec}

This work utilises data obtained by the Global Oscillation Network Group (GONG) program, managed by the National Solar Observatory, which is operated by AURA, Inc. under a cooperative agreement with the National Science Foundation. 
The data were acquired by instruments operated by the Big Bear Solar Observatory, High Altitude Observatory, Learmonth Solar Observatory, Udaipur Solar Observatory, Instituto de Astrofísica de Canarias, and Cerro Tololo Interamerican Observatory.
We thank our colleagues at the NASA Goddard Space Flight Center for contributions to this effort.
Funding for this work has been provided through the UK's STFC under grants ST/N000714 and ST/N000781.
P.F.W. is supported through the award of a Royal Astronomical Society Fellowship.

%%%%%%%%%%%%%%%%
%%%  Bibliography
%%%%%%%%%%%%%%%%

\bibliographystyle{apj}
\bibliography{RBS_big_bib}

\end{document}